# A Fully-Automatic Framework for Parkinson's Disease Diagnosis by Multi-Modality Images

Jiahang Xu[1], Fangyang Jiao[2], Yechong Huang[1], Xinzhe Luo[1], Qian Xu[3], Ling Li[4], Xueling Liu[5], Chuantao Zuo[4], Ping Wu[4]* and Xiahai Zhuang[1]*

[1] School of Data Science, Fudan University, Shanghai, China

[2] Department of Nuclear Medicine, Army Medical Center of PLA, Chongqing, China

[3] Department of Nuclear Medicine, North Huashan Hospital, Fudan University, Shanghai, China

[4] PET Center, Huashan Hospital, Fudan University, Shanghai, China

[5] Department of Radiology, Huashan Hospital, Fudan University, Shanghai, China

* Corresponding authors: Xiahai Zhuang, zxh@fudan.edu.cn
                         Ping Wu, wupingpet@fudan.edu.cn

## Abstract

**Background:** Parkinson's disease (PD) is a prevalent long-term neurodegenerative disease. Though the diagnostic criteria of PD are relatively well defined, the current medical imaging diagnostic procedures are expertise-demanding, and thus call for a higher-integrated AI-based diagnostic algorithm.

**Methods:** In this paper, we proposed an automatic, end-to-end, multi-modality diagnosis framework, including segmentation, registration, feature generation and machine learning, to process the information of the striatum for the diagnosis of PD. Multiple modalities, including T1-weighted MRI and $^{11}$C-CFT PET, were used in the proposed framework. The reliability of this framework was then validated on a dataset from the PET center of Huashan Hospital, as the dataset contains paired T1-MRI and CFT-PET images of 18 Normal (NL) subjects and 49 PD subjects.

**Results:** We obtained an accuracy of 100% for the PD/NL classification task, besides, we conducted several comparative experiments to validate the diagnosis ability of our framework.

**Conclusion:** Through experiment we illustrate that (1) automatic segmentation has the same classification effect as the manual segmentation, (2) the multi-modality images generates a better prediction than single modality images, and (3) volume feature is shown to be irrelevant to PD diagnosis.

## 1. Introduction

Parkinson's Disease (PD) is the second-to-most prevalent long-term neurodegenerative disease, as its damage to motor system mainly includes bradykinesia, rigidity and rest tremor [1]. Causing about 340,600 deaths per year, PD is one of the major concerns in neurology [2].

For clinicians, the gold standard of PD diagnostic criteria is the Movement Disorder Society Clinical Diagnostic Criteria for Parkinson's disease (MDS-PD Criteria) [1]. The MDS-PD Criteria refers to both movement disorders and nonmotor features and is made up of four supportive criteria, nine absolute exclusion criteria and ten red-flags. Subjects, meeting at least two supportive criteria in the absence of absolute exclusion criteria and without red-flag, are diagnosed with clinically established PD.

In the previous studies, the motor symptoms of PD are thought to be linked with the loss of dopaminergic neurons, and several regions of the interest (ROIs) are commonly researched, including the caudate nucleus, the putamen and the globus pallidus[3]. Therefore, the functional neuroimaging of the presynaptic dopaminergic system is underlined in the MDS-PD criteria [4]. Accordingly, several PET tracers like $^{11}$C-CFT are developed to observe the activity of dopamine transporter (DAT) [5, 6], a biomarker of presynaptic dopaminergic system which has high sensitivity in detecting early stage of PD [7].

As stated above, functional neuroimaging of the DAT like CFT-PET takes an important role in PD diagnosis. However, the information that CFT-PET alone can give is limited, for PET images are fuzzy. Therefore, the structural neuroimaging methods like T1-weighted MRI are introduced to the multi-modality diagnosis of PD [8], of which a common application is to help place the ROIs of CFT-PET. For example, Lu et al., worked on the subtypes of multiple system atrophy (MSA) utilizing T1-MRI and CFT-PET [9], while Huang et al., combined these two modalities with $^{18}$F-Fludeoxyglucose-PET (FDG-PET) and analyzed REM sleep behavior disorder [10]. In both of their studies, T1-MRI images were registered to CFT-PET images to place ROIs on the CFT-PET images.

Except for the high sensitivity, the traditional neuroimaging diagnostic procedures of PD is also known for expertise-demanding. Hence, clinicians take attempts to improve the accuracies and reduce the time consumed in diagnostic methods with the help of machine learning, among which the support volume machines (SVM) have been widely used. Long et al., used SVM to distinguish early PD patients form normal controls exploiting resting-state functional MRI [11], and obtained an accuracy of 86.96%. Haller et al., used a dataset of DTI and $^{123}$I ioflupane SPECT, and by SVM, they reached an accuracy of 97% in classifying PD from other Parkinsonism diseases [12]. Taking together, lots of works using different modalities have proved the reliability of AI-assist PD diagnosis, while few works are done based on CFT-PET.

In this work, we proposed an automatic, end-to-end, multi-modality diagnosis framework for PD, taking T1-MRI and CFT-PET images as input. Firstly, MRI images were segmented by a U-Net, from which the segmentation labels were then obtained. Secondly, ROIs of the PET images were placed according to the segmentation labels of corresponding MRI images by registration. Finally, the features were extracted from these ROIs and machine-learning-based classifier gives predictions of PD. Our main contributions are listed below:
(1) We proposed a fully-automatic framework for multi-modality PD diagnosis and tested the performance of it.
(2) We compared the U-Net segmentation with the manual segmentation in the proposed framework.

(3) We showed that MRI images help better determining the ROIs of PET.
(4) We showed the volume feature is irrelevant to PD diagnosis.

The following sections are organized as listed below. Section 2 describes the T1-MRI and CFT-PET datasets. Section 3 demonstrates the automatic diagnosis framework we proposed. Section 4 shows the experiments we conducted and the experimental details we determined to improve the performance Section 5 presents our discussion and conclusion about this work.

## 2. Material

PET images were performed by a Siemens Biograph 64 PET/CT scanner (Siemens, Munich, Germany) in three-dimensional (3D) mode. A CT transmission scan was first performed for attenuation correction. Static emission data was acquired 60 minutes after the intravenous injection of 370 MBq of $^{11}$C-CFT and lasted for 15 minutes. All subjects lay comfortably in a supine position in a room with dimmed lighting and low background noise during the scanning procedure [9, 10]. The synchronous MR imaging data were acquired using a T1-weighted 3D inversion recovery spoiled gradient recalled acquisition (IR-SPGR) with the following parameters: TE/TR = 2.8/6.6 ms, inversion time = 400 ms, flip angle = 15°, matrix = 256 × 256 × 170, field-of-view = 24 cm, slice thickness = 1 mm. Each MR image was visually inspected to rule out movement artifacts [9, 10]. MR and PET images for each subject had time interval of no more than 3 months.

Forty-nine patients with PD and eighteen age-matched normal control subjects were recruited. All subjects were screened and clinically examined by a senior investigator of movement disorders before entering the study and were followed up for at least one year. The diagnosis of PD was made with reference to the Movement Disorder Society clinical diagnostic criteria. The UPDRS motor rating and Hoehn and Yahr scale (H&Y) were assessed after cessation of oral anti-parkinsonian medications (if used) for at least 12 hours. The following exclusion criteria were used for the normal controls' recruitment: (1) tested positive by RBD Single-Question Screen [13], (2) a history of neurological or psychiatric illness, (3) prior exposure to neuroleptic agents or drug use, (4) an abnormal neurological examination. The data were summarized in **Table 1.**

All data were collected from Department of Neurology, Huashan Hospital, Fudan University, and the study was approved by the Ethics Committee of Huashan Hospital. All subjects or a legally responsible relative were given written informed consent before the study.

Table 1. Summary for the studied dataset.

| Subject | HY | Count | Gender (M/F) | Age | UPDRS |
|---------|----|----|----|----|----|
| NL | 0 | 18 | 4 / 14 | 64.1 ± 6.7 | -- |
| PD | 1 | 15 | 10 / 5 | 61.2 ± 7.6 | 14.3 ± 5.1 |
|  | 2 | 26 | 16 / 10 | 62.0 ± 7.9 | 21.6 ± 7.5 |
|  | 3 | 8 | 4 / 4 | 58.8 ± 5.9 | 34.6 ± 7.4 |

Note: for age and UPDRS, the number means mean ± standard deviation

## 3. Methodology

In this section, we demonstrated the procedures in our proposed framework, as shown in **Figure 1**. To realize the fully automatic multi-modality diagnosis of PD, three major steps were designed, as listed below:

(1) Segmentation using a U-Net

   To obtain the fine structure of the brain tissues instantly, we implemented a 3D U-Net [14] to segment the T1-MRI images.

(2) Combining Two Modalities by Registration

   After the segmentation labels of T1-MRI images were obtained, we placed ROIs on CFT-PET images by registering the anatomical MRI to CFT-PET images.

(3) Feature Extraction and Prediction

   Firstly, the caudate and putamen were segmented into subregions by k-means clustering algorithm. Secondly, we extracted features of uptakes and volumes from the ROIs. Finally, the machine-learning-based classifier gived predictions of PD.

### 3.1 Segmentation using U-Net

The first step of the proposed framework is the segmentation of MRI images, as it delineates the exact structure of striatum in the MRI images. In the subsequent steps of the framework, the segmentation was used as a reference to place the ROIs.

Conventionally, the accurate results of segmentation are obtained by manual segmentation, which is however expertise-demanding and time-consuming. The deep-learning-based segmentation has achieved state-of-the-art performance with much less time. In this work, we implemented a 3D U-Net [14,15], a state-of-the-art CNN based automatic segmentation method, to generate segmentation labels of the striatum. Different from the conventional U-Net, our network further incorporated the idea of deep supervision introduced by Mehta et al. [16] for faster training convergence and a well-designed loss function for accurate segmentation.

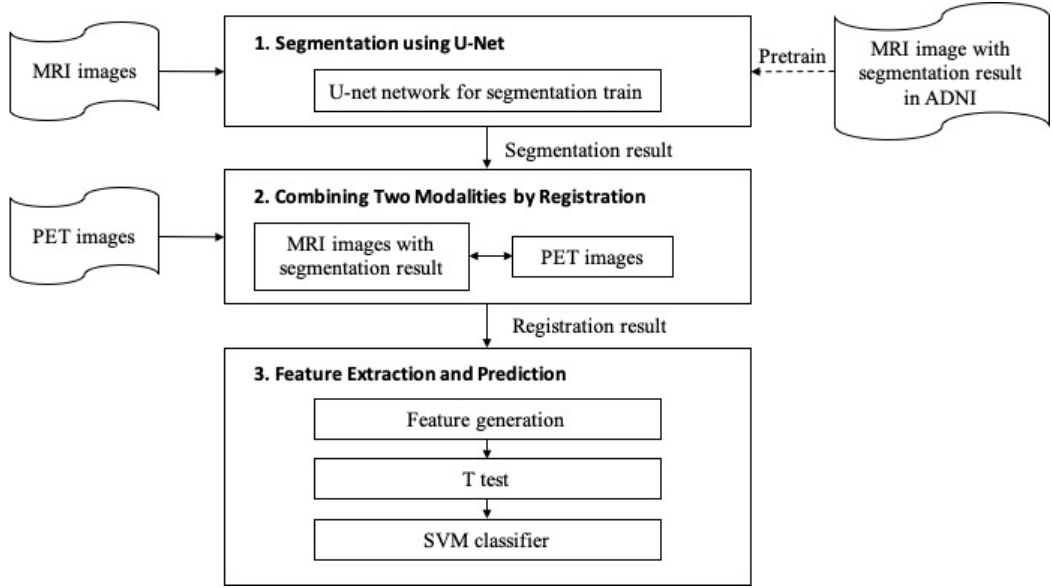

**Figure 1** The architecture of our proposed framework.

## 3.2 Combining Two Modalities Using Registration

As stated above, CFT-PET reflects the activity of DAT in the brain but have a relatively low spatial resolution. For manual diagnosis, the information that CFT-PET images give is adequate to judge the function of presynaptic dopaminergic system. However, automatic diagnosis methods require well-defined ROIs to generate features from images.

There are several ways to place well-defined ROIs of a CFT-PET image, such as referring to a pre-labeled template, and referring to a structural neuroimaging modality of same person at same scene. In our proposed method, we placed ROIs according to the segmentation labels of T1-MRI. Besides, we implemented ROI-placing according to the pre-labeled template in the comparative experiment, which will be shown in section 4. In both implementations, registration was used for propagating definition of the ROIs.

Image registration is a method which geometrically aligns two images [17]. In our framework, T1-MRI images, or the PET template, were registered to CFT-PET images to get the transformation parameters of the registration. After that, the segmentation labels were transformed to the coordinates of CFT-PET images by the transformation parameters obtained. Finally, the transformed segmentation labels were used to place ROIs over CFT-PET images. All images registrations were processed on the zxhproj platform[18], which is available on the website http://www.sdspeople.fudan.edu.cn/zhuangxiahai/0/zxhproj/. Rigid registration was used when registering T1-MRI to the corresponding CFT-PET, while rigid registration followed by affine registration was used when registering the PET template to CFT-PET images. Segmentation labels of T1-MRI images or the PET template were then transformed to the coordinates of the CFT-PET images in which way the caudate nucleus, putamen and globus pallidus, and parieto-occipital regions were labeled to be the ROIs.

### 3.3 Feature Extraction and Prediction

To extract sufficient features from the ROIs, we further divided caudate nucleus and the putamen into more substructures using a k-means algorithm [19], as each of these regions was partitioned to three clusters. After clustering, statistics of image intensity were calculated to represent the feature information in each region, including maximum, minimum, mode, $1^{st}$ and $3^{rd}$ quantile, and mean value of PET intensity. And the striatal-to-occipital ratio (SOR) was calculated for every intensity value, defined as (striatum-occipital)/occipital. Besides, the volume of the 6 anatomical ROIs were added into feature set. In all, 90 features were generated.

After features extraction, a machine-learning-based classifier was trained to classify the subjects. Here, we used Support Vector Machine (SVM), a widely used supervised learning algorithm in PD diagnosis [11, 12]. Leave one out validation strategy was used, in which for each subject in testing, the rest subjects in the dataset were used to train the model. In addition, we implemented random forest algorithm for comparisons [20], which is a widely used ensemble learning method in medical image analysis. We calculated the importance of the features by random forest, as the feature that has higher reference is more important.

## 4. Experiments

In this section, we evaluated our framework by several experiments organized by the order of the framework. In Section 4.1, the U-Net segmentation was compared with the manual segmentation to validate the reliability of the automatic segmentation. In Section 4.2, the accuracy of multi-modality images was compared with the single modality of PET images, to investigate the influence of combining multi-modality images. In Section 4.3, we discussed the influence of volumes feature of ROIs, and compared the striatal-to-occipital ratio (SOR) and the standard uptake value (SUV) with/without extracting occipital as an intensity baseline.

### 4.1 Evaluation of Automatic Segmentation for PD Diagnosis

To determine the accuracy of U-Net segmentation and validate the reliability of the segmentation step, we conducted an automatic versus manual segmentation experiment. The manual segmentation result was checked manually by a second expert to estimate inter-expert-observer variability. Both experts come from Huashan Hospital. The following steps were generated to construct the network.

#### 4.1.1 Network Architecture

**Figure 2** demonstrates our proposed network architecture for direct 3D segmentation, which is based on the U-Net [14,15] and the idea of deep supervision for fast training convergence and a well-designed loss function for accurate segmentation[16]. Specifically, the network comprises encoding and decoding paths. The encoding path is responsible for capturing contextual information by residual blocks and max-pooling operations at different resolution while the decoding path gradually recovers the spatial resolution and object boundaries. Each residual block consists of $k$ cascading $3 \times 3 \times 3$ convolutional layers of $n$ feature maps, along with batch normalization (BN), rectified linear units (ReLU) and a shortcut additional branch composed of $1 \times 1 \times 1$ convolution +

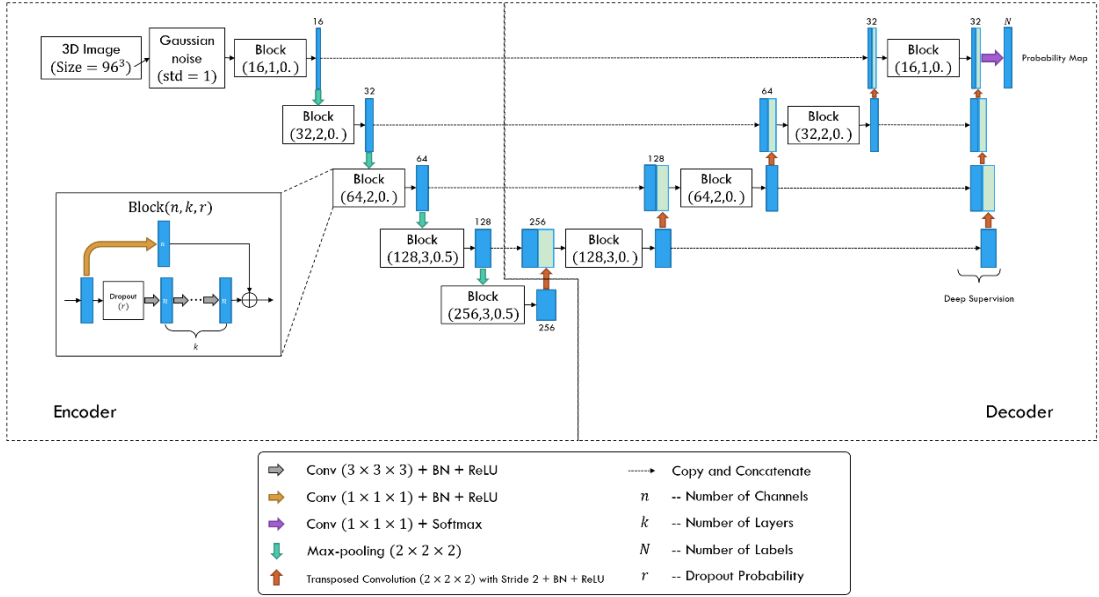

**Figure 2** The proposed segmentation network architecture.

BN + ReLU. Apart from residual blocks, the decoder applies 2 × 2 × 2 transposed convolutions with stride 2 to the output feature maps for upsampling. Besides, skip connections between the upsampled feature maps and the corresponding features of the same spatial resolution from the encoder for better convergence. Moreover, we employed the deep supervision scheme to allow more direct back-propagation to the hidden layers for faster convergence. A final 1 × 1 × 1 convolution layer with softmax function produces the segmentation probabilities. Gaussian blurring over inputs and dropout modules with rate 0.5 are also adopted to avoid overfitting.

We defined a loss function to tackle the relatively small anatomical structures of labels, as follows,
$$L = w_D L_{Dice} + w_C L_{Cross},$$
where $L_{Dice}$ denotes the exponential logarithmic Dice loss given by
$$L_{Dice} = \mathbb{E}_i[(-\ln Dice_i)^\gamma]$$
with
$$Dice_i = \frac{2(\sum_x \delta_{il}(x) \cdot p_i(x)) + \epsilon}{\sum_x (\delta_{il}(x) + p_i(x)) + \epsilon},$$
and $L_{Cross}$ denotes the cross-entropy given by
$$L_{Cross} = \mathbb{E}_x[-\ln p_l(x)].$$
Here $i$ is the segmentation label and $l$ is the ground-truth label, both at the voxel position $x$. $\delta_{il}(x)$ is the Kronecker delta, which equal 1 if $i = l$ and 0 otherwise. $p_i(x)$ is the probability of voxel $x$ being labelled as $i$.

### 4.1.2 Training Details

The segmentation U-Net were pre-trained using the Alzheimer's Disease Neuroimaging Initiative (ADNI) data with segmentation labels from the Multi-Atlas Label Propagation with Expectation-Maximization (MALPEM) platform. ADNI database collects data including magnetic resonance imaging (MRI) images as predictors of the disease, to measure and track the progression of early Alzheimer's disease (AD), which is publicly available on the website http://adni.loni.usc.edu;

MALPEM is a software package to perform whole-brain segmentation of T1-weighted MRI images, available on the website https://biomedia.doc.ic.ac.uk/software/malp-em/ [21]. In this work, 1859 T1-MRI images with corresponding segmentation result were obtained.

From the description above, the Caudate, Pallidum, Putamen area were located to be the region of interest (ROI) in this work. In addition, due to the limitation of GPU RAM and computation ability in network, we cropped ROI as big as 96 × 96 × 96 voxels of MRI images, which contains the whole structure of ROIs.

We set $w_D = 0.8$, $w_C = 0.2$ and $\gamma = 0.3$ for the loss function and trained the model on ADNI dataset using Adam optimizer with learning rate of $1 \times 10^{-3}$ for 10 epochs. Then we fine-tuned the pre-trained model using the training dataset from Huashan Hospital for another 5 epochs to obtain the desired 3D U-Net. The training dataset is a subset from the three-fold cross-validation strategy, which is elaborated on in next section.

**4.1.3 Result comparison**

With the propose of comparing the automatic segmentation results with the gold standard using three-fold cross-validation, we split the whole dataset into three disjoint folds and fine-tuned the model on the union of every two disjoint subsets, which provided three different trained models. After testing the performance on the corresponding complementary subsets, we finally calculated the Dice coefficients of each subject by assuming that the three models were homogeneous in making predictions.

The visualization of the five segmentation cases produced by the proposed network is provided in **Figure 3**. It is observed that structures of some anatomy are either underestimated or overestimated with respect to their volumes and boundaries, which may affect the subsequent prediction accuracy.

The average Dice coefficients of each anatomy are illustrated in **Table 2**. The left pallidum (colored as goldenrod in **Figure 3**) is worst segmented and with the most inter-subject variation while the right putamen (colored as olive drab in **Figure 3**) is best segmented and with the least inter-subject variation.

Then, we used both segmentation results to complete the follow-up steps of our framework. The accuracy (acc) was computed for the leave-one-out cross-validation in the machine learning module. For this work, the accuracy of the two experiment both reached 100% accuracy on PD/NL task, indicating no difference between the automatic or manual segmentation from T1-MRI for ROI definition and feature extraction for PD diagnosis. In next section, we studied the difference of ROI definition without T1-MRI for single modality PD diagnosis.

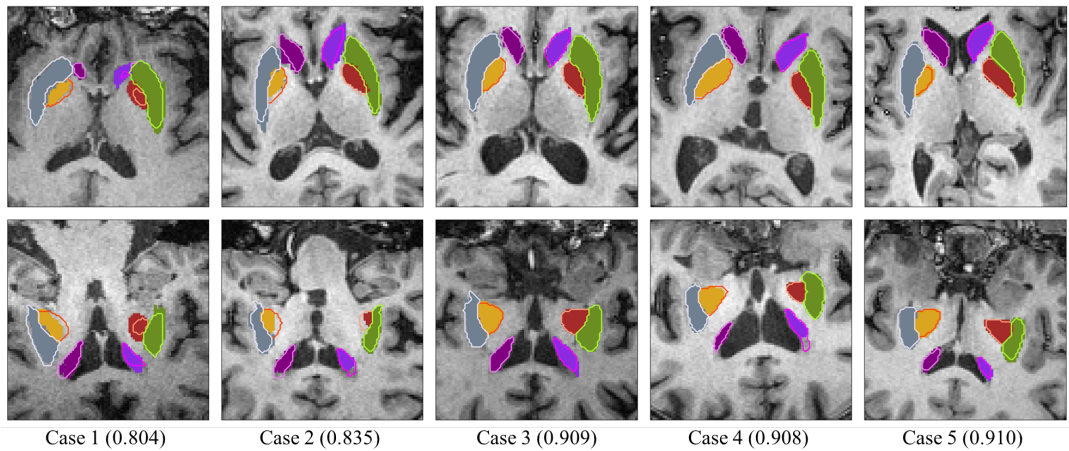

| Case 1 (0.804) | Case 2 (0.835) | Case 3 (0.909) | Case 4 (0.908) | Case 5 (0.910) |

**Figure 3** Visualization of the segmentation results, slices of the axial view (top row) and the coronal view (bottom row). The colored blocks and contours represent the ground truths masks and the automatic segmentation boundaries, respectively. Case 1 and case 2 are two worst segmentations, and case 3, case 4 and case 5 are three median results. Values in the parentheses refer to the corresponding DSCs.

**Table 2** Average DSCs of the segmentation of each anatomy with their corresponding inter-subject variations.

|  | Right Caudate | Left Caudate | Right Pallidum | Left Pallidum | Right Putamen | Left Putamen |
|---|---|---|---|---|---|---|
| Dice ± inter-subject variation (%) (95% confidence interval) | 88.5 ± 6.3 | 90.1 ± 7.2 | 89.3 ± 11.4 | 86.9 ± 13.0 | 92.2 ± 5.0 | 91.4 ± 5.5 |

## 4.2 Influence of Multi-Modality Images

The following experiments were designed to evaluate the effects of multi-modality images and single modality PET on PD diagnosis. In the experimental group of multi-modality images, MRI provided anatomical structures and obtained the segmentation results of the region of interest containing precise boundary in the PET image by registration. In the control group of single PET images, the PET images were registered to a pre-labeled AAL PET template, and the same affine matrix were used to obtain the segmentation result of PET. Then, we conducted the rest of the experiment in the same way for the two group.

In the comparative experiment, the accuracy of multi-modality model reached 100% on PD/NL task, and the accuracy of single PET model was 98.51% on the same task. Furthermore, we trained the classifier using features of multi-modality method and tested it using features of single PET method. The accuracy dropped to 88.05%, namely 8 subjects were misclassified.

Table 3 Feature importance of method with/without volume feature.

| feature name | Importance With Volume Feature | Importance Without Volume Feature |
|---|---|---|
| mean | 0.20295131 | 0.20060408 |
| med | 0.20240259 | 0.19645433 |
| 3$^{rd}$ quantile | 0.19605019 | 0.19460838 |
| 1$^{st}$ quantile | 0.18249546 | 0.18346047 |
| max | 0.14429148 | 0.14974968 |
| min | 0.0715383 | 0.07512307 |
| volume | 0.00027068 | -- |

### 4.3 Feature Selection

For features generated from intensity of striatum were well-known to be helpful on PD diagnosis, we mainly discussed two features, volume of brain structure and the radioactive uptake of the parieto-occipital region in this section.

### 4.3.1 Influence of Volumes

To verify roles of volume, we compared the experimental results of adding volume as features with the results without adding volume factors. In order to obtain an accurate volume, the volume of the MRI image without downsampling was chosen to generate the volume feature. We calculated the number of voxels occupied by the labels of ROIs and converted them into a same world coordinate to prevent the influence of different spacing on MRI images. The final volume feature was normalized for SVM model fitting.

In the comparative experiment, the accuracy with/without volume feature were both reached 100%. The feature importance of the two model is shown in **Table 3**, from which we can find the importance of volume is negligible.

### 4.3.2 Influence of the Baseline

Several works characterized radioactive uptake by the striatal-to-occipital ratio (SOR), defined as (striatum-occipital)/occipital, where striatum and occipital represent the intensity value in the PET images, as the parieto-occipital region is widely considered of lacking DAT distribution [22-24]. To verify the influence of the parieto-occipital region as a baseline, the following experiments were designed.

We generated the feature of parieto-occipital region uptake in two ways. One is using the PET intensity of the parieto-occipital region as a baseline and calculate the striatal-to-occipital ratio (SOR) for all intensity-related feature, and the other is regarding the feature of parieto-occipital region uptake as a new feature for the SVM classifier directly. Together with the feature generation without the intensity of parieto-occipital region, called standard uptake value (SUV), three models were compared to show the influence of it.

**Table 4** P values of t-test results from the selected features.

| ROI | | SOR | | | | | | volme |
|---|---|---|---|---|---|---|---|---|
| | | mean | max | min | 1st quantile | median | 3rd quantile | |
| Right Caudate | front | 6.52E-10 | 2.55E-09 | 4.65E-06 | 2.87E-09 | 1.84E-09 | 3.41E-10 | 0.0544 |
| | middle | 7.02E-10 | 4.76E-10 | 0.001585 | 6.32E-09 | 7.02E-10 | 2.46E-10 | |
| | rear | 1.04E-07 | 1.78E-09 | 2.21E-07 | 3.94E-06 | 6.00E-07 | 3.06E-08 | |
| Left Caudate | front | 1.11E-08 | 2.70E-08 | 3.20E-07 | 2.75E-08 | 2.11E-08 | 8.46E-09 | 0.0321 |
| | middle | 5.52E-08 | 4.72E-08 | 2.51E-05 | 1.65E-07 | 6.76E-08 | 4.74E-08 | |
| | rear | 1.41E-06 | 2.03E-07 | 0.000103 | 3.78E-06 | 3.45E-06 | 1.82E-06 | |
| Right Putamen | front | 7.37E-18 | 2.80E-16 | 3.16E-07 | 2.03E-17 | 1.64E-17 | 1.19E-17 | 0.0469 |
| | middle | 1.18E-30 | 7.62E-26 | 1.13E-09 | 5.21E-14 | 6.58E-31 | 1.79E-30 | |
| | rear | 1.03E-13 | 1.57E-13 | 1.02E-07 | 1.20E-12 | 1.50E-13 | 6.86E-14 | |
| Left Putamen | front | 9.50E-15 | 5.80E-15 | 2.11E-06 | 5.24E-14 | 3.66E-14 | 8.96E-15 | 0.02 |
| | middle | 1.01E-27 | 3.03E-22 | 2.17E-10 | 1.80E-28 | 2.36E-28 | 5.97E-27 | |
| | rear | 3.26E-31 | 8.34E-14 | 9.98E-11 | 3.18E-29 | 1.02E-30 | 9.95E-32 | |
| Right Pallidum | | 8.84E-23 | 4.07E-21 | 7.53E-07 | 4.72E-16 | 1.27E-21 | 1.60E-12 | 0.002 |
| Left Pallidum | | 6.06E-19 | 1.50E-17 | 4.87E-07 | 1.75E-07 | 2.90E-08 | 2.13E-09 | 0.309 |

The accuracy of the three model were all 100%, indicating the characteristic of the parieto-occipital region has no significant effect on this task.

## 5. Discussion and Conclusion

In this work, we proposed a fully automatic framework for PD diagnosis. This method exploits two modalities, CFT-PET and T1-MRI, and it contains three major steps: segmentation, registration, feature extraction and prediction. A total of 90 features were selected, including the statistics of the radioactive uptake ratios and the volume information for each region. A t-test was performed to evaluate the significance of every feature, with result shown in **Table 4**.

To validate the performance of the proposed method, we implemented it on the PD/NL dataset. For segmentation, the framework used a U-Net that pre-trained with the ADNI dataset and fine-tuned with the Huashan dataset, to segment the striatum of the MRI images. In the registration step, the T1-MRI images were registered to CFT-PET images. The ROIs of CFT-PET images were then identified according to the segmentation labels. Finally, the framework extracted the features and gave predictions by a machine-learning-based classifier. Consequently, the framework reached 100% accuracy for the PD/NL task, which shows its ability to diagnose PD combining CFT-PET and T1-MRI.

In the proposed method, one of the main differences to the traditional methods lies in that the ROIs are placed according to the labels of the automatic segmentation. To evaluate the performance of the U-Net, we calculated the Dice coefficients between automatic and manual segmentation labels. Besides, we implemented a variant of the proposed method, which places ROIs according to the manual segmentation method. As a result, this variant also reached 100% accuracy for the PD/NL task. Hence, both manual and automatic segmentation can well place ROIs for the PD/NL task.

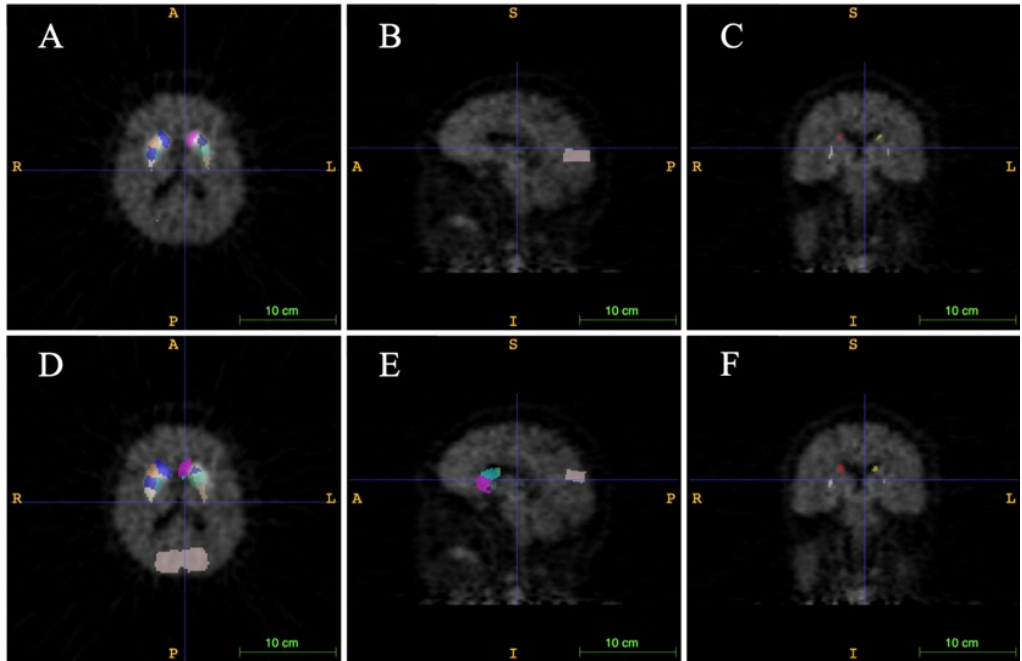

**Figure 4** The comparison of gold standard and wrongly placed ROIs of the wrongly predicted subject. A, B and C show the segmentation result in gold standard, D, E and F show the segmentation in the wrongly predicted subject. Images in the same column have same cursor position, as the first column is horizontal plane, the second column is sagittal plane, and the third column is coronal plane.

Besides, we tried the other way of placing ROIs, which doesn't need structural medical imaging modality and place ROIs according to a pre-segmented PET template, to prove that anatomical information helps the proposed framework better diagnose AD. In this implementation, AAL PET template was used as the PET template, and was registered to CFT-PET images. The rest steps of this implementation were the same as the proposed multi-modality framework. This implementation reached 98.51% accuracy on the same PD/NL task, being inferior to the proposed multi-modality framework. The result indicates that though using single modality CFT-PET is enough to give a favorable prediction of PD for AI-based diagnosis, the multi-modality framework has a better performance, as the structural information of the T1-MRI images helps better placing ROIs.

To investigate the inferior performance of the single-modality method, we visualize the wrongly predicted subject in **Figure 4**. One can see that the main problem lies in the erroneous delineation of the ROIs, which is the key step in feature extraction for the learning-based prediction algorithm. This error comes from the fact that defining ROIs according to a PET template ignores the unique brain structures of different subjects, and the precision of ROIs can greatly influence the accuracy of prediction.

Furthermore, we tested the uniformity of the classifiers of different implementations by training a classifier using features of multi-modality manual segmentation method and testing it using features of other methods. When testing with features of multi-modality automatic segmentation method, we still obtained 100% accuracy indicating that the features of manual and automatic segmentation are highly consistent. However, testing with single modality method had an accuracy of 88.05%. This lower accuracy implies that the features of single modality method may not be well extracted due to the wrongly placed ROIs. Hence, when comparing with multi-modality methods, single modality method by naturally needs more feature engineering and better designed AI algorithms.

Since this framework has been established, we evaluated the importance of the extracted features. **Figure 5** shows the importance of different categories of variables in the gold standard experiment using manual segmentation results, our automated segmentation results experiments, and experiments without volume feature. The most important variables are mean and quantile statistics, then is maximum. Minimum has a lower importance than the first five features, and the variable with the lowest importance is volume. This could be reason that the volume of striatum does not change significantly with the progression of PD, as concluded from **Table 4** and the literature [8]. The reason why minimum is less important may due to the wrongly placed ROIs, for the striatum region has a higher SUV, and a small offset of ROIs can result in a huge bias.

**Figure 6** shows the heatmap of feature distribution displaying the influence of each subregions for the classification in PD/NL task. The pattern of difference is expressed according to the color scales. From **Figure 6**, one can find the most relevant region influencing the separation of PD/NL are localized in the middle and rear of putamen, then pallidum, and the caudate reveal the least significance on this task.

Moreover, serval future works can be done based on the current pipeline. Firstly, the classifiers can be trained with Parkinsonian disorders (PDS) dataset to classify PD and atypical PDS like Multiple System Atrophy (MSA) and Progressive Supranuclear Palsy (PSP), as this task has increasingly important clinical values. Secondly, only medical imaging information is involved in this framework, while other aspects of information, including age, gender, motor ratings and other biomarkers can be considered. Thirdly, the quantity of subjects in this work is relatively small, and a bigger dataset can be used to both validate and improve the performance of the framework. Finally, more works can be done to apply the state-of-the-art algorithms to the feature extraction or classification step.

To conclude, we proposed a fully automatic framework, combining two modalities, T1-MRI and CFT-PET, for PD diagnosis. This framework has been trained and tested by the dataset and reached 100% accuracy on the PD/NL task. To ensure the performance of the framework, we used multi-modality method, and trained a U-Net to segment T1-MRI images. This work also emphasizes the high reference value the CFT-PET holds in the PD diagnosis.

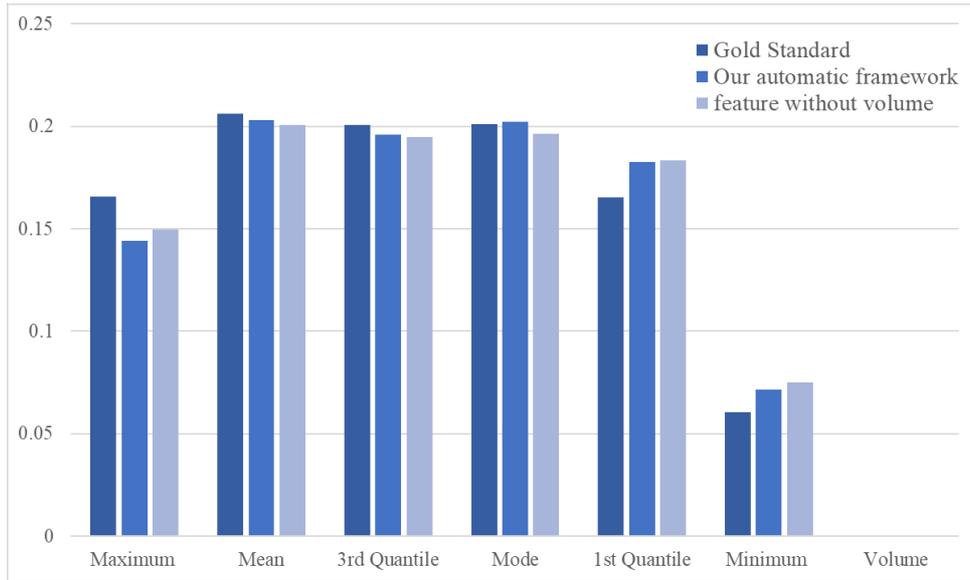

**Figure 5** The importance of different categories in different methods

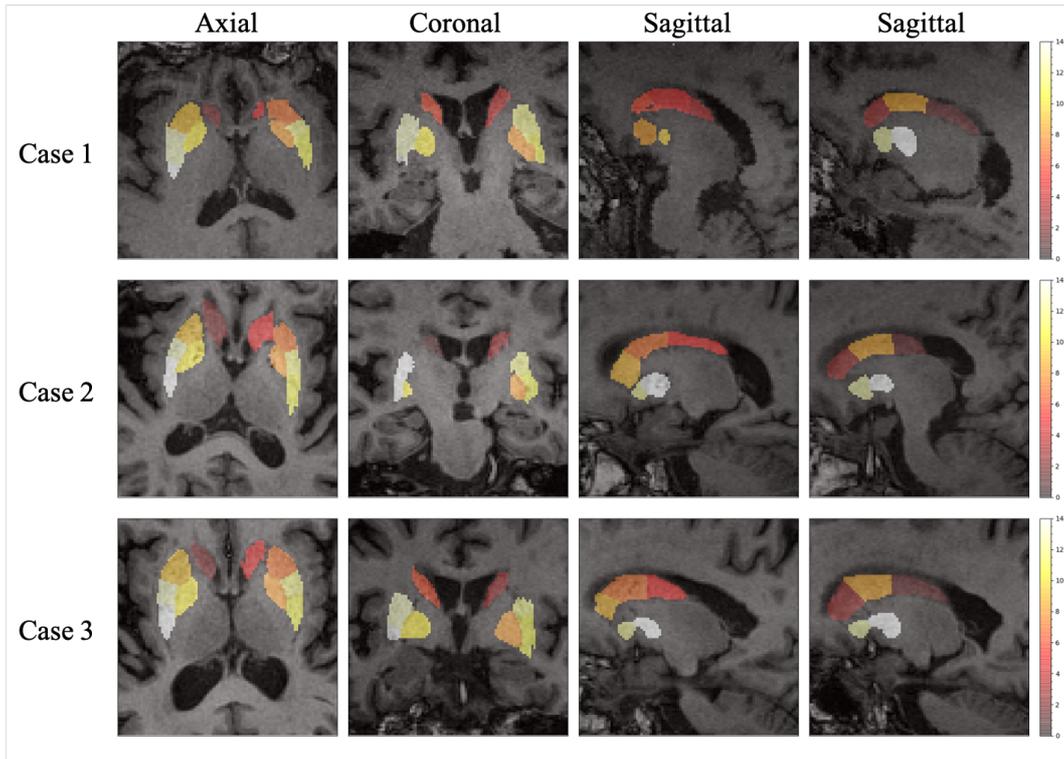

**Figure 6** The importance of ROIs in the proposed framework. One axial slice, one coronal slice and two sagittal slices of three subjects are chosen to show the importance heatmap of the ROIs.